\documentclass[
    ,final            % use final for the camera ready runs
  ]
  {aipproc}

\layoutstyle{8x11double}

\usepackage{amsmath}	% required for `\align' (yatex added)
\newcommand{\phibar}{\overline{\phi}}

\begin{document}

\title{Observing dynamical SUSY breaking with lattice simulation}

\classification{
%<Replace this text with PACS numbers; choose from this list:
%                \texttt{http://www.aip..org/pacs/index.html}>
%11.15.Ha Lattice gauge theory
%11.30.Pb Supersymmetry 
%11.30.Qc Spontaneous and radiative symmetry breaking
11.15.Ha, 11.30.Pb, 11.30.Qc
}

\author{Issaku Kanamori}{
  address={Theoretical Physics Laboratory,
RIKEN, Wako,
%2-1 Hirosawa, Wako
Saitama 351-0198, JAPAN }
}

\begin{abstract}
On the basis of the recently developed lattice formulation of supersymmetric
theories which keeps a part of the supersymmetry, we propose 
a method of observing dynamical SUSY breaking with lattice simulation.
We use Hamiltonian as an order parameter and measure the ground state
energy as a zero temperature limit of the finite temperature simulation.
Our method provides a way of obtaining a physical result from the
lattice simulation for supersymmetric theories.
\end{abstract}

\maketitle

%%%%%%%%%%%%%%%%%%%%%%%%%%%%%%%%%%%%%%%%%%%%
%% MAINMATTER
%%%%%%%%%%%%%%%%%%%%%%%%%%%%%%%%%%%%%%%%%%%%

\section{Introduction}
In these several years, lattice formulation of supersymmetric theories
made a great progress \cite{Kaplan:2002wv, Sugino:2003yb,
Catterall:2004np,D'Adda:2004jb,Suzuki:2005dx, Elliott:2005bd} 
(for the review, see \cite{Giedt:2007hz}).
Not only the formulations but the simulations have been already started
\cite{Catterall:2006jw, Suzuki:2007jt, Fukaya:2007ci}.

We have developed a method of observing dynamical SUSY breaking using
a lattice simulation \cite{prd, ptp}.
Since the SUSY is not broken in the perturbation if it is not broken in the
tree level, a way of observing SUSY breaking due to non-perturbative
effects is very important.
Usually, the Witten index provides such a method.
But sometimes it is not available.  
In fact the main target in this talk,
two-dimensional $\mathcal{N}=(2,2)$ super Yang-Mills theory (SYM)
is such an example.  
What we know is only a conjecture which states it may be broken
\cite{Hori:2006dk}.

Our method uses the Hamiltonian as an order parameter.
The requirement for the lattice model is an exact $Q$-symmetry 
on the lattice so it can be widely used in principle.
After a very quick explanation of putting SUSY on the
lattice using Sugino model \cite{Sugino:2003yb}, we state our method of 
measuring the Hamiltonian.
After confirming that the method actually works, we will show
the result for the super Yang-Mills case.

\section{Lattice Model}

We use a lattice model with one exactly kept supercharge.
It seems impossible to put the SUSY on the lattice, because
SUSY algebra contains infinitesimal translation but on the lattice
we have only finite translations.
However, what we have realized in the recent development is that
it is possible if $\mathcal{N}\geq 2$, especially in low-dimensional case.
Most of the recent development utilize the topological twist.
After the twist, we have a scalar supercharge instead of spinors.
We can put the scalar on a lattice site and keep it exact at finite
lattice spacing.
Some of the simulation have already done aiming the check
of the formulation.

We adopt such a model proposed by Sugino \cite{Sugino:2003yb}.
The target theory in the continuum has twisted $\mathcal{N}=2$ supersymmetry.
It has the following algebra:
\begin{align}
   Q^2 &= \delta_\phi^{\rm (gauge)}, \qquad
   Q_0^2 = \delta_{\phibar}^{\rm (gauge)} ,\\
 \{Q, Q_0\} &= 2i\partial_0 +2\delta_{A_0}^{\rm (gauge)},
\label{eq:algebra}
\end{align}
where we pick up only a part of the whole algebra.
$\delta_{\ \bullet}^{\rm (gauge)}$ denotes infinitesimal gauge
transformation with the parameter $\bullet$.
After the twist, we have four supercharges, a scalar $Q$,
two from a two-dimensional vector $Q_\mu$ ($\mu=0,1$), and a pseudo
scalar $\tilde{Q}$.
They are nilpotent up to gauge transformation.
The field components are a complex scalar field $\phi$
($\phibar=\phi^\dagger$),  gauge field $A_\mu$,
Majorana fermions in the twisted basis $\eta$, $\chi$ and $\psi_\mu$,
and an auxiliary field $H$.

The point is that the action has $Q$-exact form $S=Q\Lambda$,
where $\Lambda$ is a gauge invariant quantity, and $Q$
can be discretized keeping the nilpotency even at a finite lattice spacing.
Therefore, the lattice action defined as $Q$-exact quantity is
manifestly $Q$-invariant and enjoys $Q$-symmetry at a finite lattice
spacing as long as it is gauge invariant.
The lattice version of the $Q$ transformation are the following:
\begin{align*}
   Q U(x,\mu) &= i\psi_\mu(x) U(x,\mu), \\
   Q \psi_\mu(x)&=i\psi_\mu(x)\psi_\mu \\
  & \lefteqn{\quad -i\bigl(\phi(x)-U(x,\mu)\phi(x+\hat{\mu}\bigr)U(x,\mu)^{-1}),} \\
   Q \phi(x) &= 0,\\
 Q \chi(x) &= H(X),
& Q H(x) &= [\phi(x), \chi(x)], \\
 Q \phibar(x) &= \eta(x),
& Q \eta(x) &= [\phi(x), \phibar(x)],
\end{align*}
where $U(x,\mu)$ is a gauge link variable.

Eventually, we have nilpotent $Q$ and $Q$-invariant action.
The remaining $Q_0,\ Q_1$ and $\tilde{Q}$ will be automatically restored
in the continuum limit.

\section{Method}
We use Hamiltonian as the order parameter since 
it is zero if and only if the SUSY is not broken.
Therefore, it is very sensitive about the choice of the origin of the
Hamiltonian.  
We use the SUSY algebra to define the Hamiltonian.
From the algebra (\ref{eq:algebra}), we have
\begin{align}
 Q \mathcal{J}_0^{(0)}&= 2 \mathcal{H},
\end{align}
where $\mathcal{J}_0^{(0)}$ is the 0-th component of the 
Noether current for $Q_0$ and $\mathcal{H}$ is a Hamiltonian density.
On the lattice we have only $Q$ transformation but no $Q_0$
transformation.
Therefore, we discretize the continuum version of the Noether current by
hand.
%This is the discretized version of $\mathcal{J}_0^{(0)}$.
We know the $Q$ transformation so it is straightforward to obtain the 
Hamiltonian.

An advantage of measuring the Hamiltonian is that it is a one-point
function.  Compared with two-point functions which are used for
measuring spectrum, it is numerically very cheap and easy to 
calculate.

Another point of our method is the boundary condition.
In the lattice simulation, the lattice size is limited and usually we
use a periodic condition.
For the current purpose, however, we need a different
choice.

As usual method for observing spontaneous symmetry breaking,
we apply an external field conjugate to the order parameter.
The conjugate to the Hamiltonian is the temperature.
That is, we impose the \emph{anti-periodic} condition in the time direction
for fermion.
Therefore we break SUSY by boundary condition or equivalently by the 
temperature.
Then we take zero temperature limit and observe the effect of the 
breaking is left or not.

It is interesting to see what would happen if we took the periodic
conditions.  Under the periodic condition, the simulation does not
work.  
Since the periodic partition function is exactly the Witten index,
it is easy to see the expectation value of the Hamiltonian is
proportional to the derivative of the Witten index 
$\langle \exp(-\beta H)\rangle_{\rm PBC}$:
\begin{align}
 \langle H \rangle_{\rm PBC} 
 &=\frac{-\frac{\partial}{\partial \beta} \langle \exp(-\beta H)\rangle_{\rm
 PBC}}{\langle \exp(-\beta H)\rangle_{\rm PBC}}.
\end{align} 
Because the index does not depend on the coupling $\beta$\footnote{
In the anti-periodic case, $\beta$ can be understood as the inverse temperature.
}, the derivative w.r.t. $\beta$ is zero so the expectation value of the
Hamiltonian seems zero as well.  If the SUSY is broken, however, 
the denominator is also zero and the expectation value of the
Hamiltonian can be non-zero value.
In the simulation, what we measure is the numerator.
The path integral is replace by an ensemble average and the normalization
factor is replaced by the size of the ensemble. 
%Furthermore, under the periodic boundary condition, 
%expectation values of $Q$-exact quantity is always zero. 
Therefore, the simulation should give always zero even if the SUSY is broken
under the periodic boundary condition.

In fact the expectation value of our $Q$-exact
Hamiltonian is always zero under the periodic boundary condition, 
if we ignore the possible zero of the denominator.  
Since the action is $Q$-invariant and the $Q$ is
nilpotent, the expectation value of 
$Q$-exact quantity is zero.  This is consistent with
the above argument.

Eventually, we should measure the ground state energy as a
zero temperature limit of the $Q$-exact Hamiltonian.

\section{Result}

\begin{figure}
 \includegraphics[width=.9\linewidth,height=.65\linewidth]{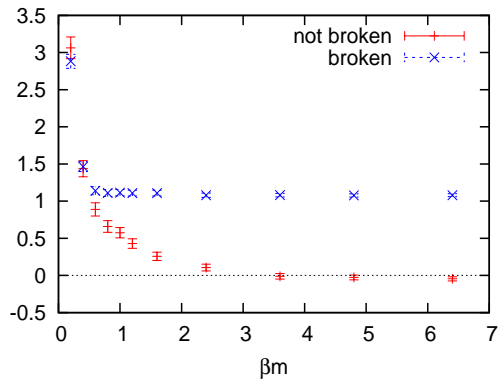}
 \label{fig:sqm}
 \caption{
Hamiltonian of SQM versus inverse
 temperature $\beta$, in unit a dimensionful parameter $m$ in the potential.  
The plot is after taking the continuum limit.
The  ``not broken'' data goes to zero as the 
inverse temperature $\beta$ goes large,
while the ``broken'' data stays finite.  We can easily distinguish these 
two cases so our method actually works.
}
\end{figure}

As a check of our method, we first investigate Supersymmetric Quantum
Mechanics (SQM).
The known fact is that the form of the potential decides whether 
SUSY is broken or not.
The lattice model we use was given in \cite{Catterall:2000rv} and keeps
nilpotent $Q$ and $Q$-exact action.
The figure \ref{fig:sqm} shows that our method actually works.
\begin{figure}
 \includegraphics[width=.9\linewidth,height=.61\linewidth]{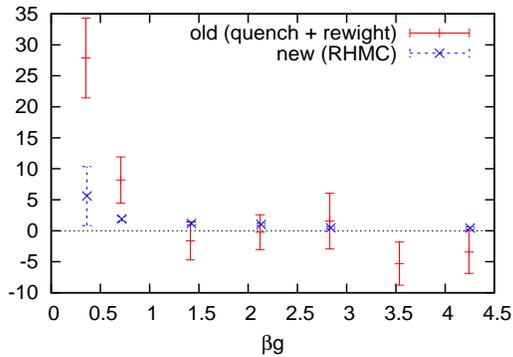}
 \label{fig:SYM}
 \caption{Hamiltonian of SYM versus inverse temperature $\beta$, in unit of the 
dimensionful coupling $g$.  The old data was obtained by reweighting
 method and the new data was obtained by RHMC method.
The both indicate small ground state energy.
}
\end{figure}
\begin{figure}
 \includegraphics[width=.9\linewidth,height=.61\linewidth]{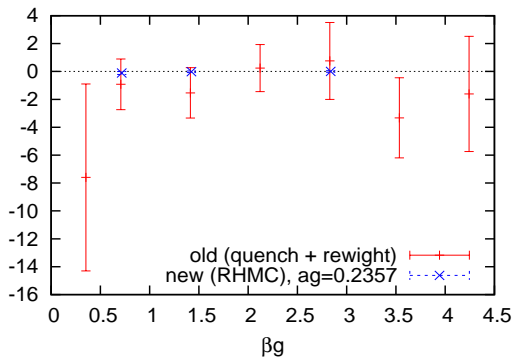}
 \label{fig:pbcSYM}
 \caption{Hamiltonian versus inverse temperature $\beta$, under the \emph{
 periodic} boundary condition.  
 In the reweighting method the continuum limit is taken while in the
 RHMC method the lattice spacing is fixed at $ag=0.2357$.
 The values should be zero even at the finite lattice spacing.
}
\end{figure}
Next let us investigate the SYM case.  See figure \ref{fig:SYM}.
We plot results obtained by two different simulation algorithms,
new one and old one presented in \cite{prd, ptp}.
The errors are drastically reduced in the new plot and the ground state
energy seems small.\footnote{
Before giving conclusive statement on the SUSY breaking,
we should confirm that the lattice model we use actually describes the
target theory in the continuum \cite{restore}.
}
Because of the $Q$-exactness, the periodic Hamiltonian 
should be zero.  In fact it is consistent with zero within the error (figure
\ref{fig:pbcSYM}) and consistent with
the discussion based on the Witten index.

%detail
Some details of the simulation of SYM are in order.
The fermion effect is treated by a reweighting method or Rational Hybrid
Monte Carlo (RHMC).
The reweighting method uses quenched configurations and treats the 
fermion effect as a part of the observable so it is some how indirect.  
The RHMC method uses
configurations with fermion effects. 
The simulation codes are developed based on the FermiQCD
\cite{DiPierro:2000bd} 
and parameters for rational approximation in the RHMC are obtained from
\cite{remez}.
We fix the physical spacial size to $1.41$ in unit of dimensionful gauge
coupling $g$.\footnote{All the dimensionful quantities are 
measured in unit of $g$.}
The parameters are the following.
The lattice size is $3\times 6$--$36\times 12$ and the lattice spacing is
$0.0707$--$0.2357$.  The number of the independent configurations are
$9900$--$99900$ in the reweighting method and $10$--$1700$ in the RHMC
method.

\section{Conclusion}
We have developed a method of observing dynamical SUSY breaking using
lattice simulation.
Our method is available if the lattice model %of interest
has one exactly 
kept supercharge $Q$ and $Q$-exact action.
We used Hamiltonian as the order parameter and measure the ground state 
energy.
It actually worked in supersymmetric quantum
mechanics, a system whether SUSY is broken or not was already known.
Then we applied it to two-dimensional super Yang-Mills. 
It is the first physical application of the recent 
development of the lattice SUSY. 
The application to the other system is straightforward.
Now we can use the lattice formulation for supersymmetric theories.

%%%%%%%%%%%%%%%%%%%%%%%%%%%%%%%%%%%%%%%%%%%%%%%%
%% BACKMATTER
%%%%%%%%%%%%%%%%%%%%%%%%%%%%%%%%%%%%%%%%%%%%%%%%

\begin{theacknowledgments}
The author would like to thank M.~Hanada, H.~Kawai, H.~Matsufuru, K.~Murakami, 
F.~Sugino and H.~Suzuki.
He is supported by the Special Postdoctoral Researchers Program 
at RIKEN.  
The simulation result of super Yang-Mill model was obtained by using
Riken Super Combined Cluster System (RSCC).
\end{theacknowledgments}




%\bibliography{sample}

\IfFileExists{\jobname.bbl}{}
 {\typeout{}
  \typeout{******************************************}
  \typeout{** Please run "bibtex \jobname" to optain}
  \typeout{** the bibliography and then re-run LaTeX}
  \typeout{** twice to fix the references!}
  \typeout{******************************************}
  \typeout{}
 }

%%%%%%%%%%%%%%%%%%%%%%%%%%%%%%%%%%%%%%%%%%%
%% The following lines show an example how to produce a bibliography
%% without the help of the BibTeX program. This could be used instead
%% of the above.
%%%%%%%%%%%%%%%%%%%%%%%%%%%%%%%%%%%%%%%%%%%

\end{document}